%% file: paper.tex
\documentclass[twocolumn,showpacs,aps,prl,superscriptaddress]{revtex4}

\usepackage{graphicx}
\usepackage{dcolumn}
\usepackage{amsmath}
\usepackage{epsfig}
\usepackage{dcolumn}

\newcolumntype{.}{D{.}{.}{-1}}
\newcolumntype{,}[1]{D{.}{.}{#1}}
\newcolumntype{p}{D{\%}{\%}{3}}
\newcolumntype{a}{D{.}{\to}{-1}}

\input pubboard/babarsym.tex

\newcommand{\BABARPubYear}    {05}
\newcommand{\BABARPubNumber}  {047}

\newcommand{\SLACPubNumber} {11520}

\def\figurebox#1#2#3{%
    \def\arg{#3}%
    \ifx\arg\empty
    {\hfill\vbox{\hsize#2\hrule\hbox to #2{\vrule\hfill\vbox to #1{\hsize#2\vfill}\vrule}\hrule}\hfill}%
    \else
    {\hfill\epsfbox{#3}\hfill}%
    \fi}

\newcommand{\Dsm}      {\ensuremath{D^-_s}\xspace}
\newcommand{\Dssm}     {\ensuremath{D^{*-}_s}\xspace}
\newcommand{\DsMultm}  {\ensuremath{D^{(*)-}_s}\xspace}
\newcommand {\vecp} {\ensuremath{\kern 0.2em\vec{\kern 0.1em p}{}\xspace}}

\begin{document}

\preprint{\babar-PUB-\BABARPubYear/\BABARPubNumber} 
\preprint{SLAC-PUB-\SLACPubNumber} 

\begin{flushleft}
\babar-PUB-\BABARPubYear/\BABARPubNumber\\
SLAC-PUB-\SLACPubNumber\\
\end{flushleft}

\title{
{\large \bf
Search for rare quark-annihilation decays, {\boldmath $\Bub \to \DsMultm \phi$}} 
}

\input pubboard/authors_sep2005.tex

\begin{abstract}
\input abstract.tex

\end{abstract}

\pacs{13.25.Hw, 12.60.Jv, 11.30.Pb}

\maketitle

\begin{figure}[htb]
\begin{center}
\epsfig{file=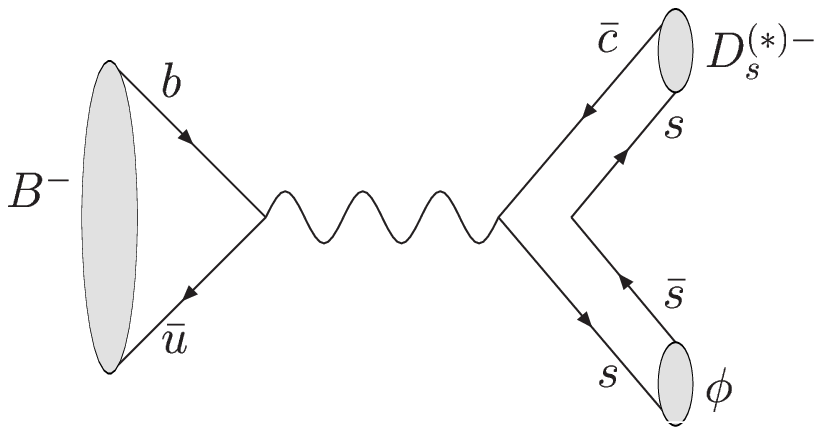,height=4.5cm}
\caption{Feynman diagram for $\Bub \to D_s^{(*)-} \phi$.}
\label{fig:feyn}
\end{center}
\end{figure}

In the Standard Model (SM), the decay $\Bub \to \DsMultm \phi$ 
occurs through annihilation of the two quarks in the
$B$ meson into a virtual $W$ as shown in Figure~\ref{fig:feyn}.
No pure $W$-annihilation $B$ decays have ever been observed.
The current upper limits on the branching fractions of
$\Bub \to \Dsm \phi$ and 
$\Bub \to \Dssm \phi$ are $3.2 \times 10^{-4}$ (90\% C.L.) and 
$4 \times 10^{-4}$ (90\% C.L.), respectively, set by the 
CLEO collaboration in 1993~\cite{cleo}.

In the SM, $B$ annihilation amplitudes are highly suppressed.  Calculations
of the $\Bub \to \Dsm \phi$ branching fraction give predictions of $3
\times 10^{-7}$ using a perturbative QCD approach~\cite{lu}, $1.9
\times 10^{-6}$ using factorization~\cite{mohanta}, and $7 \times
10^{-7}$ using QCD-improved factorization~\cite{mohanta}.

Since the current experimental limits are about three orders of
magnitude higher than the SM expectations, searches for $\Bub \to
\DsMultm \phi$ could be sensitive to new physics contributions. 
Reference~\cite{mohanta} argues that the branching fraction for
$\Bub \to \Dsm \phi$ could be as high as $8 \times 10^{-6}$ in a
two-Higgs doublet model and $3 \times 10^{-4}$ in the minimal
supersymmetric model with $R$-parity violation, depending on
the details of the new physics parameters.


Our results are based on $234 \times 10^6$ $\FourS\to B\Bbar$ decays,
corresponding to an integrated luminosity of 212 fb$^{-1}$, collected
between 1999 and 2004 with the \babar\ detector~\cite{babar} at the 
\pep2\  \BF~\cite{pep2} at the Stanford Linear Accelerator Center.
A 12~fb$^{-1}$ off-resonance data sample,
with a center-of-mass (CM) energy 40~\mev below the \FourS resonance peak,
is used to study continuum events, $e^+ e^- \to q \bar{q}$
($q=u,d,s,$ or $c$).
The number of $B$ mesons in our data
sample is two orders of magnitudes larger than in the
previously-published search~\cite{cleo}.


We search for the decay
$\Bub \to \DsMultm \phi$
in the following modes:
$\Dssm \to \Dsm \gamma$,
$\Dsm \to \phi \pi^-$, $\KS K^-$, and $K^{*0}K^-$, with the secondary 
decay modes 
$\phi \to K^+ K^-$, $\KS \to \pi^+ \pi^-$, and
$K^{*0} \to K^+ \pi^-$.
(Charge-conjugate decay modes are implied throughout this article.)
We denote the $\phi$ from $\Bub \to \DsMultm \phi$
as the ``bachelor $\phi$'' in order to distinguish it
from the $\phi$ in $\Dsm \to \phi \pi^-$.

Our analysis strategy is the following. Starting from the set of
reconstructed tracks and electromagnetic calorimeter clusters, we
select events which are kinematically consistent with the \upsbb, 
$\Bub\to\DsMultm\phi$ hypothesis. 
Backgrounds, mostly from continuum events, are suppressed using a
likelihood constructed from a number of kinematical and event shape
variables. For each candidate satisfying all selection criteria, 
we calculate the energy-substituted mass, \mes, defined later in this article.
The \mes distribution of these events is then fit to a signal plus 
background hypothesis to extract the final signal yield.

All kaon candidate tracks in the reconstructed decay chains must
satisfy a set of loose kaon identification criteria based on the
response of the internally-reflecting ring-imaging Cherenkov radiation
detector and the ionization measurements in the drift chamber
and the silicon vertex tracker. The kaon
selection efficiency is a function of momentum and polar angle, and is
typically 95\%.  These requirements provide a rejection factor of
order 10 against pion backgrounds.  No particle identification
requirements are imposed on pion candidate tracks.

We select $\phi$, $\KS$, and $K^{*0}$ candidates from pairs of
oppositely-charged tracks with invariant masses consistent with the
parent particle decay hypothesis and consistent with originating from
a common vertex.  The invariant mass requirements are $\pm 10$~\mev
($\sim 2.4 \Gamma$) for the $\phi$, $\pm 9$~\mev ($\sim 3 \sigma$) for
the $\KS$, and $\pm 75$~\mev ($\sim 1.5 \Gamma$) for the $K^{*0}$, where
$\sigma$ and $\Gamma$ are the experimental and natural width, 
respectively, of these particles.  
(Here, and throughout the paper, we use natural units where $c=1$). We
then form $D_s^-$ candidates in the three modes listed above by
combining $\phi$, $\KS$, or $K^{*0}$ candidates with an additional
track.  The invariant mass of the $D_s^-$ candidate must be within 15~\mev
 ($\sim 3\sigma$) of the known $D_s^-$ mass.  In the $D_s^- \to
\phi \pi^-$ and $D_s^- \to K^{*0}K^-$ modes, all three charged tracks
are required to originate from a common vertex.  In the $D_s^- \to \KS
K^-$ mode, the $\KS$ and $D_s^-$ vertices are required to be separated
by at least 3 mm.  This last requirement is very effective in
rejecting combinatorial background and is 94\% efficient for signal.
We select $\Dssm$ candidates from $D_s^-$ and photon candidates.
The photon candidates are constructed from calorimeter clusters with
lateral profiles consistent with photon showers and with energy above
60~\mev in the laboratory frame.  We require that the mass difference
$\Delta M \equiv M(\Dssm) - M(\Dsm)$ be between 130
and 156~\mev.  The $\Delta M$ resolution is about 5~\mev.  

At each stage in the reconstruction, the measurement of the
momentum vector of an intermediate particle is improved by 
refitting the momenta of the decay products with kinematical
constraints.  These constraints are based on the known mass~\cite{PDG}
of the intermediate particle and on the fact that the 
decay products must originate from a common point in space.  

Finally, we select \Bm candidates by combining \DsMultm and
bachelor $\phi$ candidates. A \Bm candidate is characterized
kinematically by the energy-substituted mass 
$\mes \equiv
\sqrt{(\frac{1}{2} s + \vecp_0\cdot \vecp_B)^2/E_0^2 - \vecp_B^2}$ 
and energy difference $\Delta E \equiv E_B^*-\frac{1}{2}\sqrt{s}$,
where $E$ and $\vecp$ are energy and momentum, the asterisk denotes the CM
frame, the subscripts $0$ and $B$ refer to the initial \FourS and $B$
candidate, respectively, and $s$ is the square of the CM energy.  In
the CM frame, $\mes$ reduces to $\mes = \sqrt{\frac{1}{4}s -
\vecp^{*2}_B}$.  For signal events we expect $\mes \sim M_B$, the known
$\Bub$ mass, and $\Delta E \sim 0$.  The resolutions of $\mes$ and
$\Delta E$ are approximately 2.6~\mev and 10~\mev, respectively.  For
events with more than one $\Bub$ candidate, we retain the
candidate with the lowest $\chi^2$
computed from the measured values, known values, and resolutions for
the $D_s^-$ mass, the bachelor $\phi$ mass, and, where applicable,
$\Delta M$.

This analysis was performed blind: GEANT4 simulated data~\cite{geant}
or data samples outside the fit region were used for background studies
and selection criteria optimization.  
Most of the backgrounds to the $\Bub \to \DsMultm
\phi$ signal were determined to be from continuum events.  
To reduce these backgrounds we make two additional requirements.
First, we require $|\cos\theta_T| < 0.9$, where $\theta_T$ is the
angle between the thrust axis of the $\Bub$ candidate and the rest of
the tracks and neutral clusters in the event, calculated in the CM
frame.  The distribution of $|\cos\theta_T|$ is essentially uniform
for signal events and strongly peaked near unity for continuum events.
Second, for each event we define a relative likelihood for signal and
background based on a number of kinematical quantities.  The relative
likelihood is defined as the ratio of the likelihoods for signal and
background.  The signal (background) likelihood is defined as the
product of the probability density functions, PDFs, for the various
kinematical quantities in signal (background) events.

The kinematical quantities used in the likelihood are 
reconstructed masses, helicity angles, and a Fisher
discriminant designed to distinguish between 
continuum and $B\Bbar$ events; we will discuss each of these 
in turn in the following paragraphs. All PDFs
are chosen based on studies of Monte Carlo and
off-resonance data.

The masses used in the likelihoods are those of the \Dsm, $\Delta M$
for $D_s^{*-} \to D_s \gamma$, the $K^{*0}$ in $D_s^- \to K^{*0} K^-$, and
the $\phi$ in $D_s^- \to \phi\pi^-$.  The signal PDFs
for the mass variables are the sum of two Gaussian distributions for
\Dsm and $\Delta M$, a Breit-Wigner distribution for the $K^{*0}$, and
a Voigtian distribution~\cite{voigtian} for the $\phi$.  We
parameterize the background PDFs as uniform distributions.  Note that
the mass of the bachelor $\phi$ and the mass of the $\KS$ in $D_s^-
\to \KS K^-$ are not used in the definition of the likelihoods.  This
is because studies of background event samples suggest that background
events contain mostly real bachelor $\phi$ and real $\KS$ mesons.

The helicity angles used in the likelihood are those 
in $K^{*0} \to K^+ \pi^-$ and in $\phi \to K^+K^-$
for both $\phi$ candidates.
The signal PDFs for these quantities are set by angular
momentum conservation to be proportional to $\cos^2\theta$,
where $\theta$ is the helicity angle for the process.
The one exception is the helicity angle distribution of
the bachelor $\phi$ in $\Bub \to D_s^{*-} \phi$,
where the polarization of the two vector mesons in the
final state is not known.  For this reason, the helicity
angle of the bachelor $\phi$ is not used in the definition
of the likelihood for the $\Bub \to D_s^{*-} \phi$ mode.
The background PDFs for these variables are uniform in $\cos\theta$.
In addition, in the likelihood we also use the 
polar angle of the $\Bub$ candidate in the CM frame ($\theta_B$).
The signal is expected to follow a $\sin^2\theta_B$ distribution,
while the background is independent of $\cos\theta_B$.

The final component of the likelihood is a Fisher discriminant
constructed from the quantities $L_0 = \sum_i{p^*_i}$ and $L_2 =
\sum_i{p^*_i \cos^2\alpha^*_i}$. Here, $p^*_i$ is the magnitude of the momentum and
$\alpha^*_i$ is the angle with respect to the thrust axis of the
$\Bub$ candidate of tracks and clusters not used to reconstruct the
$\Bub$, all in the CM frame.  The signal and background PDFs for this
variable are modeled as bifurcated Gaussians with different means and
standard deviations. Note that the likelihood is constructed to have
negligible correlations between its components.  Since the Fisher
discriminant is highly correlated with the $|\cos\theta_T|$ variable
defined above, the $|\cos\theta_T|$ variable is treated separately and
not included in the likelihood.

The likelihood requirements have been optimized to maximize the
expected sensitivity on a mode-by-mode basis. (The expected
sensitivity is defined as the branching ratio upper limit that we
should obtain, on average, when performing this experiment many times
in the absence of a signal.)  Depending on the mode, the combined
efficiency of the likelihood and the $|\cos\theta_T|$ requirements
varies between 71\% and 83\%, while providing a rejection factor of
between 4 and 7 against continuum backgrounds.

After applying the requirements on relative likelihood and 
$|\cos\theta_T|$, we also demand that $\Delta E$ fall inside the signal
region: within 
30~\mev ($\sim 3 \sigma$) of its expected mean value for signal events.
This mean value is determined from simulation, and varies
between $-3$ and $0\mev$, depending on the mode.

\begin{table}[bt] 
\caption{\protect
Efficiencies ($\epsilon$), branching fractions (${\cal B}$),
and products of efficiency and branching fractions
for the modes used in the
$\Bub \to \DsMultm \phi$ search. The uncertainties on 
the $\epsilon$ and ${\cal B}$ are discussed in the text.
Here ${\cal B}$ is the product of branching fractions for
the secondary and tertiary decays in the specified decay mode.}
\begin{center}
\begin{tabular}{l ,{3} . ,{5}} \hline\hline
B Mode ~~\Dsm Mode           & \multicolumn{1}{c}{$\epsilon$} 
& \multicolumn{1}{c}{~${\cal B} (10^{-3})$~} 
& \multicolumn{1}{c}{~$\epsilon \times {\cal B} (10^{-3})$} \\ \hline
\multicolumn{1}{l}{$\Bub \rightarrow D_s^- \phi$} & & & \\
~~~~~~~~~$D_s^- \to \phi \pi^-$   & ~~0.192~~ & 11.6 & 2.22\\
~~~~~~~~~$D_s^- \to K^-  \KS$   & ~~0.177~~ & 8.2 & 1.45\\
~~~~~~~~~$D_s^- \to K^{*0} K^-$   & ~~0.140~~ & 14.5 & 2.03\\
\hline
\multicolumn{1}{l}{$\Bub \to D_s^{*-} \phi$} & & & \\
~~~~~~~~~$D_s^- \to \phi\pi^-$    & ~~0.109~~ & 10.9 & 1.19\\
~~~~~~~~~$D_s^- \to K^-  \KS$   & ~~0.100~~ & 7.7 & 0.77\\
~~~~~~~~~$D_s^- \to K^{*0} K^-$   & ~~0.083~~ & 13.6 & 1.14\\ 
\hline\hline
\end{tabular}
\label{tab:accBR}
\end{center}
\end{table}

The efficiencies of our selection requirements, shown in
Table~\ref{tab:accBR}, are determined from simulations.  For
the $\Bub \to D_s^{*-}\phi$ mode, we take the average of the
efficiencies calculated assuming fully longitudinal or transverse
polarization for the two vector meson final state.  These efficiencies
are found to be the same to within 1\%.  The quantities ${\cal B}$ in
Table~\ref{tab:accBR} are the product of the known branching fractions
for the secondary and tertiary decay modes.  These are taken from the
compilation of the Particle Data Group~\cite{PDG}, with the exception
of the branching fraction for $\Dsm \to \phi \pi^-$, for
which we use the latest, most precise measurement ${\cal B}(\Dsm \to
\phi \pi^-) = (4.8 \pm 0.6)$\%~\cite{dsphipi}.  Since the branching
fractions for the other two $\Dsm$ modes are measured with respect to
the $\Dsm \to \phi \pi^-$ mode, we have rescaled their tabulated
values from the Particle Data Group accordingly.


The systematic uncertainties on the products of efficiency and
branching ratio for the secondary decays in the decay chain of
interest are summarized in Table~\ref{tab:accErr}.  The largest
systematic uncertainty is associated with the uncertainty on the $\Dsm
\to \phi \pi^-$ branching ratio, which is only known to
13\%~\cite{dsphipi}, and which is used to normalize all other $\Dsm$
branching ratios.

\begin{table}[bt]
\caption{\protect
Systematic uncertainties on
$\sum_i \epsilon_i \cdot {\cal B}_i$,
where the index $i$ runs over the three $D_s^-$ modes 
used in this analysis, $\epsilon_i$ are the experimental
efficiencies, and ${\cal B}_i$
are the branching fractions for the $i^\mathrm{th}$ mode.}
\begin{center}
\begin{tabular}{l p p} \hline \hline
Source & \multicolumn{1}{c}{$\Bub \to D_s^- \phi$}
&\multicolumn{1}{c}{$\Bub \to D_s^{*-} \phi$} \\ \hline
\Dsm branching fraction & 14\% & 14\% \\
\Dssm branching fraction & - & 2.5\% \\
Other branching fractions & 1.5\% & 1.5\% \\
Charged kaon ID & 13.2\% & 13.3\% \\
Tracking and $\KS$ efficiency & 3.7\% & 3.7\% \\
Photon efficiency & - & 1.8\% \\
Final state polarization & - & 1\% \\
Selection requirements & 5\% & 5\% \\ 
Simulation statistics & 0.6\% & 0.6\% \\
\hline
Total & 20\% & 21\% \\ 
\hline\hline
\end{tabular}
\label{tab:accErr}
\end{center}
\end{table}

The experimental systematic uncertainties all relate to the
determination of the efficiency, $\epsilon$.  The dominant source
of error is the uncertainty in the efficiency of the kaon
identification requirements.  The efficiency of these requirements is
calibrated using a sample of kinematically identified $D^{*0} \to D^0
\pi^+$, $D^{0} \to K^- \pi^+$ decays, where track quality selection differences 
between this sample and our analysis sample have been taken into
account.  The efficiency of the kaon identification requirements is
assigned a 3.6\% systematic error.
This results in a systematic uncertainty of 14\% for
the efficiency of the modes with four charged kaons ($\Dsm \to K^{*0}K^-$,
 $\Dsm \to \phi \pi^-$), and 11\% for the mode
with three charged kaons ($\Dsm \to \KS K^-$).  A second class of
uncertainties is associated with the detection efficiency for tracks
and clusters.  From studies of a variety of
control samples, the tracking efficiency is understood at the level of
1.4\% (0.6\%) for transverse momenta below (above) 200~\mev.  There
is also a 1.9\% uncertainty associated wih the reconstruction of
$\KS \to \pi^+\pi^-$ which can occur a few centimeters away from
the interaction point.  Given the multiplicity and momentum spectrum
of tracks in the decay modes of interest, the uncertainty on the
efficiency of reconstructing tracks in the $B$-decay chain is
estimated to be 3.7\%.  In the $\Bub \to D_s^{*-} \phi$ search, there is
an additional uncertainty of 1.8\% due to the uncertainty on the
efficiency to reconstruct the photon in $D_s^{*-} \to D_s \gamma$, 
and a 1\% uncertainty from the unknown polarization in the
final state.  Finally, to ascertain the systematic 
uncertainty due to the
efficiency of the other event selection requirements, we examine 
the variation of the efficiency under differing conditions:
shifting the $\Delta E$ by
3~\mev (0.3\%); shifting the mean of the \Dsm and $\phi$ masses and
$\Delta M$ by 1~\mev~ (0.2\%, 0.1\%, 0.2\%, respectively); increasing
the width of the \Dsm and $\phi$ masses and $\Delta M$ by 1~\mev~
(1.5\%, 0.4\%, 1.5\%, respectively); using a Fisher distribution
obtained from the data sample of a similar analysis, $B\to D\pi$ with
$D\to K\pi$ (3\%).  Thus we assign a 5\% systematic on the combined
efficiency of these selection criteria.


We determine the yield of signal events from an unbinned extended
maximum-likelihood fit to the $\mes$ distribution of $\Bub$ candidates
satisfying all of the requirements listed above.  
We fit simultaneously in two $|\Delta E|$ regions:
in the signal region the distribution is
parametrized as a Gaussian and the combinatorial
background as a threshold function~\cite{ARGUS}; in a sideband of
$\Delta E$ ($|\Delta E| < 200$~\mev, excluding the
signal region) we fit solely for the threshold function parameter.
In our fit, the
amplitude of the Gaussian is allowed to fluctuate to negative values,
but, for reasons of numerical stability, the sum of the Gaussian and
the threshold function is constrained to be positive over the full
$\mes$ fit range.  The mean and the standard deviations of the
Gaussian are constrained to the values determined from Monte Carlo
simulation. The fitting procedure was extensively tested with 
sets of simulated data, and was found to provide an 
unbiased estimate of the signal yield.

\begin{figure}[bt]
\begin{center}
\epsfig{file=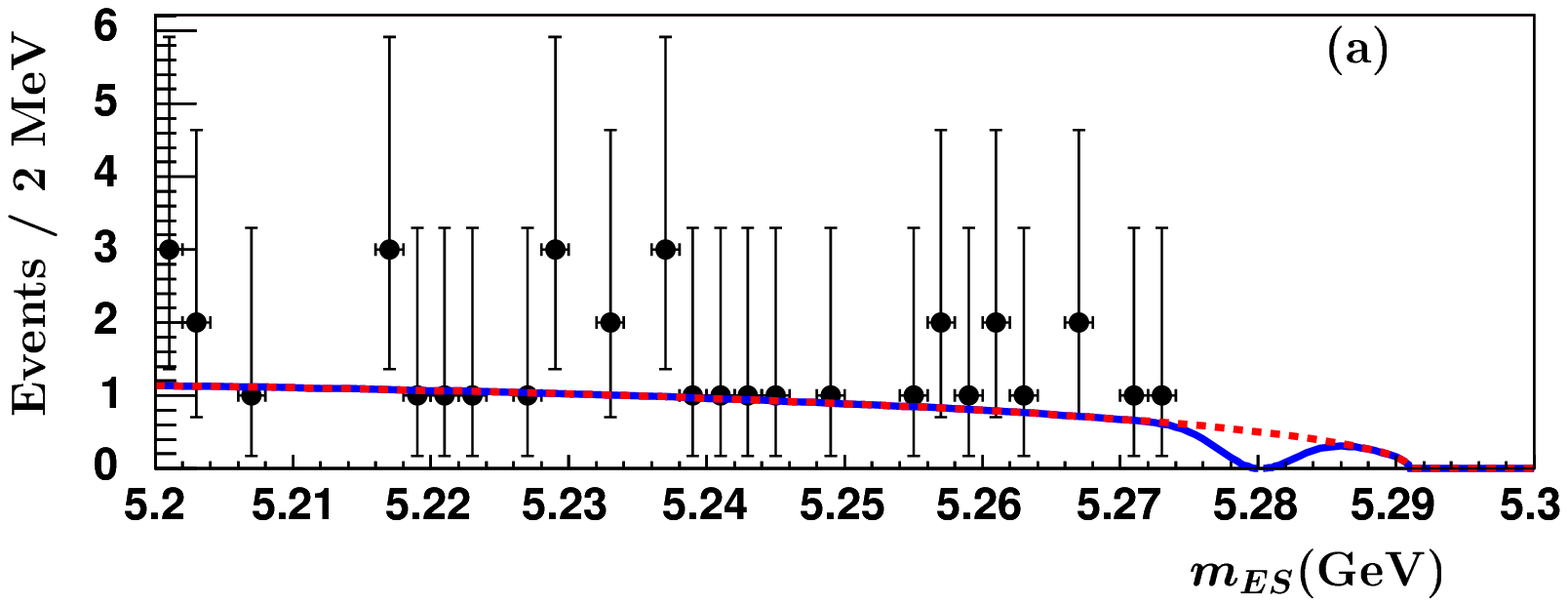,width=\linewidth}
\epsfig{file=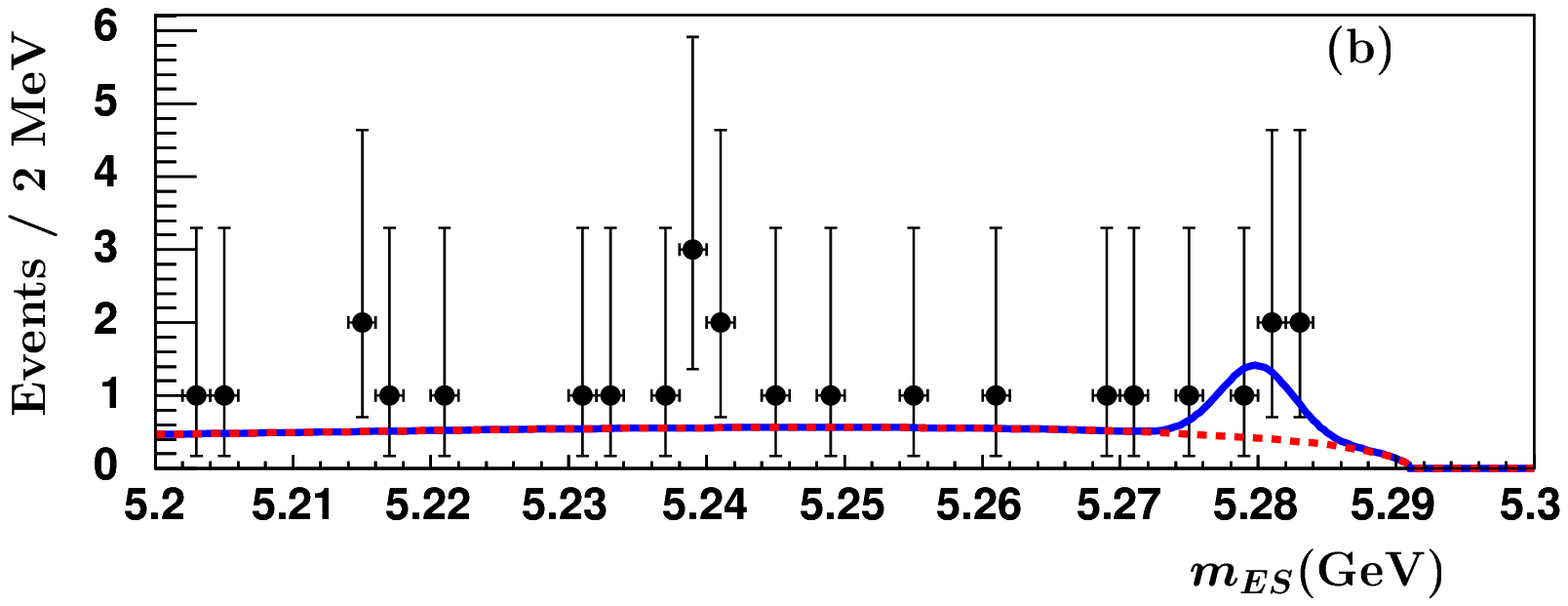,width=\linewidth}
\caption{Distribution of $\mes$ for (a) $\Bub \to D_s^- \phi$ and
(b) $\Bub \to D_s^{*-} \phi$ candidates in the $|\Delta E|$ signal region.  
The superimposed curves are
the result of the fits described in the text. The dashed curve is the 
background contribution and the solid curve is the sum of the 
signal and background components.}
\label{fig:mesfit}
\end{center}
\end{figure}

\begin{figure}[bt]
\begin{center}
\epsfig{file=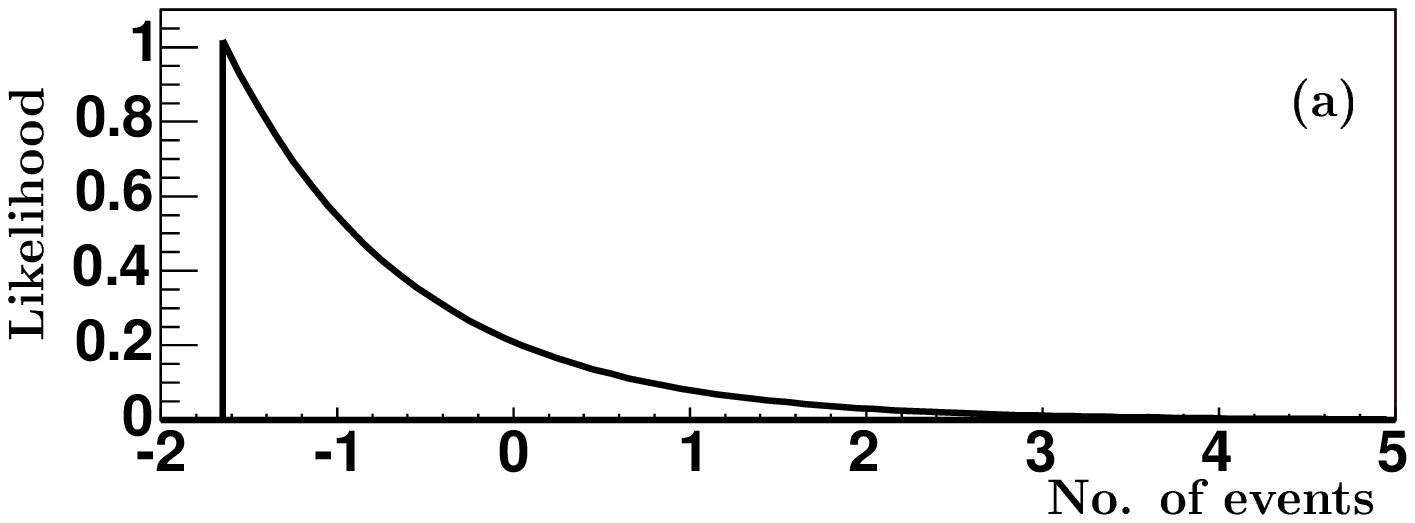,width=\linewidth}
\epsfig{file=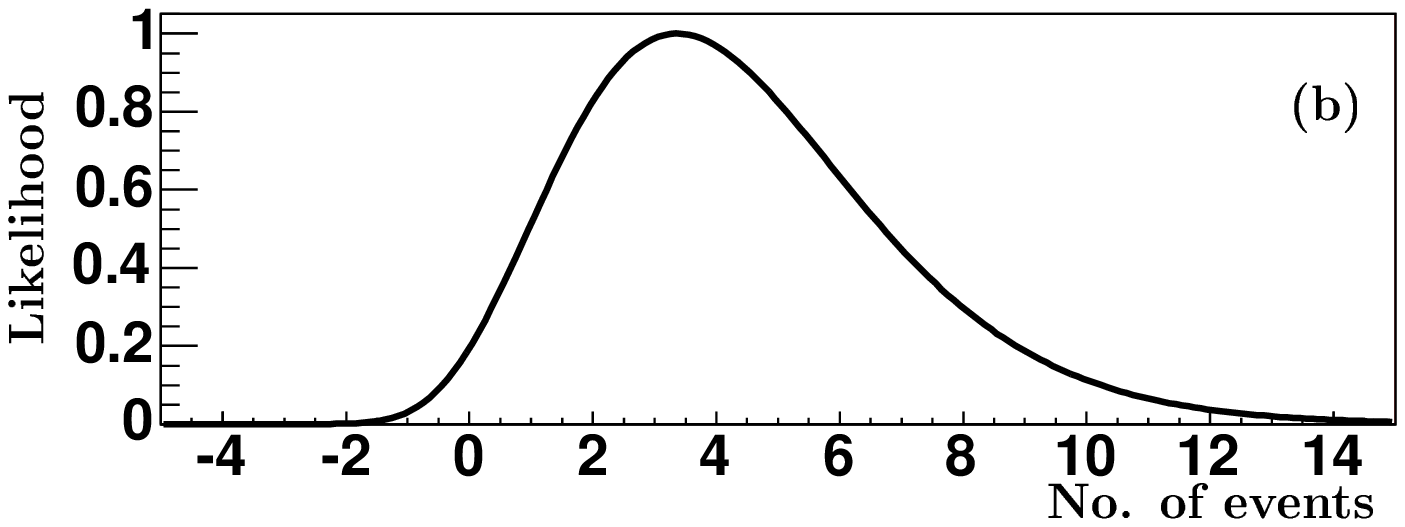,width=\linewidth}
\caption{Likelihood from the fit in arbitrary units as a 
function of the number of signal events.
(a) $\Bub\to\Dsm\phi$;
(b) $\Bub\to\Dssm\phi$.}
\label{fig:lik}
\end{center}
\end{figure}

Figure~\ref{fig:mesfit} shows the $\mes$ distribution of the selected
candidates.  We see no evidence for $\Bub \to \DsMultm \phi$.  The
fitted event yields are $N = -1.6^{+0.7}_{-0.0}$ and $N =
3.4^{+2.8}_{-2.1}$ for the $\Bub \to D_s^- \phi$ and $\Bub \to D_s^{*-}
\phi$ modes, respectively, where the quoted uncertainties correspond
to changes of 0.5 in the log-likelihood for the fit.  The
likelihood curves are shown in Figure~\ref{fig:lik}.  The requirement
that the sum of the Gaussian and the threshold function be always
positive results in an effective constraint $N > -1.6$ in the $\Bub \to
\Dsm \phi$ mode.  This is the source of the sharp edge at $N = -1.6$ in
the likelihood distribution of Figure~\ref{fig:lik}(a).

We use a Bayesian approach with a flat prior to set 90\% confidence
level upper limits on the branching fractions for the $\Bub \to D_s^-
\phi$ and $\Bub \to D_s^{*-} \phi$ modes.  In a given mode, the upper
limit on the number of observed events ($N_{UL}$) is defined as

\begin{equation}
\int_0^{N_{UL}} {\cal L}(N)~dN~~~=~~~0.9 \int_0^{+\infty}{\cal L}(N)~dN.
\end{equation}

\noindent ${\cal L}(N)$ is the likelihood, determined from the $\mes$ fit 
detailed above, as a function of the number of signal events, $N$.
The upper limit ${\cal B}$ on the branching fraction is

\begin{equation}
{\cal B} < \frac{N_{UL}}{N_{\BB} \sum_i \epsilon_i \times {\cal B}_i}.
\end{equation}

\noindent $N_{\BB} = (233.9 \pm 2.5) \times 10^6$
is the number of $B\Bbar$ events, the index $i$ runs over
the three $D_s^-$ decay modes, $\epsilon_i$ is the efficiency in the
$i^\mathrm{th}$ mode, and ${\cal B}_{i}$ is the product of all secondary and
tertiary branching fractions (see Table~\ref{tab:accBR}).

We account for systematic uncertainties by numerically convolving
${\cal L}(N)$ with a Gaussian distribution with its width determined by
the total systematic uncertainties (Table~\ref{tab:accErr}) in the two
modes, including the 1.1\% uncertainty in $N_{\BB}$ added in
quadrature.  We find limits ${\cal B}(\Bub \to D_s^- \phi) < 1.9 \times
10^{-6}$ and ${\cal B}(\Bub \to D_s^{*-} \phi) < 1.2 \times 10^{-5}$ at
the 90\% confidence level.  These limits are calculated using ${\cal
B}(\Dsm \to \phi \pi^-) = (4.8 \pm 0.6)$\% from
Reference~\cite{dsphipi}.  If we were to use the value ${\cal B}(\Dsm
\to \phi \pi^-) = (3.6 \pm 0.9)$\% from the Particle Data Group~\cite{PDG}, we
would find ${\cal B}(\Bub \to D_s^- \phi) < 2.7 \times 10^{-6}$ and
${\cal B}(\Bub \to D_s^{*-} \phi) < 1.7 \times 10^{-5}$.  For
completeness, we also compute 
${\cal B}(\Bub \to D_s^- \phi) \times {\cal B}(\Dsm \to \phi \pi^-) < 8.6 \times 10^{-8}$ 
and 
${\cal B}(\Bub\to D _s^{*-} \phi) \times {\cal B}(\Dsm \to \phi \pi^-) < 5.4 \times 10^{-7}$, 
also at the 90\% confidence level.


In summary, we have searched for $\Bub \to \DsMultm \phi$, and we
have found no evidence for these decays.  Our limits are about two
orders of magnitude lower than the previous results, but are still one
order of magnitude higher than the Standard Model expectation.  

Using the calculation of Reference~\cite{mohanta}, we can use our
results to set bounds on new physics contributions. These bounds are
obtained in the framework of factorization, neglecting systematic 
uncertainties associated with the calculation of hadronic effects. 
In the type II Two-Higgs-Doublet model we extract a tree-level
90\% C.L. limit 
$\tan\beta$/M$_{\rm H} < 0.37$/GeV, where $\tan\beta$ is the ratio of
the vacuum expectation values for the two Higgs doublets and M$_{\rm H}$
is the mass of the charged Higgs boson.  This limit is not quite
as stringent
as the limit that can be obtained from the $\Bub \to \taum \nut$ decay mode,
$\tan\beta$/M$_{\rm H} <0.29$/GeV~\cite{belletaunu}.

In the context of supersymmetric models with $R$-parity violation 
(RPV)~\cite{mohanta,susy},
the new physics contribution to $B \to D_s \phi$ depends
on the quantity
$\frac{\lambda^2}{M^2} 
\equiv \sum_i \frac{\lambda^{\prime}_{2i2} \lambda^{\prime *}_{i13}}{M^2_i}$,
where $\lambda^\prime_{jkl}$ is the coupling between the $j^\mathrm{th}$ 
generation doublet lepton superfield, the $k^\mathrm{th}$ generation doublet
quark superfield, and the $l^\mathrm{th}$ generation singlet down-type 
quark superfield; $M_i$ is the mass of the $i^\mathrm{th}$ generation
charged super-lepton.
Conservatively assuming maximal destructive interference between the
SM and RPV amplitudes,
we find $|\frac{\lambda^2}{M^2}| < 4 \times 10^{-4}/$(100 GeV)$^2$.


\input pubboard/acknow_PRL.tex

\end{document}

%% file: pubboard/authors_sep2005.tex
%
\author{B.~Aubert}
\author{R.~Barate}
\author{D.~Boutigny}
\author{F.~Couderc}
\author{Y.~Karyotakis}
\author{J.~P.~Lees}
\author{V.~Poireau}
\author{V.~Tisserand}
\author{A.~Zghiche}
\affiliation{Laboratoire de Physique des Particules, F-74941 Annecy-le-Vieux, France }
\author{E.~Grauges}
\affiliation{IFAE, Universitat Autonoma de Barcelona, E-08193 Bellaterra, Barcelona, Spain }
% ...
\collaboration{The \babar\ Collaboration}
\noaffiliation

%% file: abstract.tex
We report on searches for $B^- \to \ensuremath{D^-_s}\xspace \phi$ and
$B^- \to \ensuremath{D^{*-}_s}\xspace \phi$.  In the context of the
Standard Model, these decays are expected to be highly suppressed
since they proceed through annihilation of the $b$ and $\bar{u}$
quarks in the $B^-$ meson.  Our results are based on 234 million
$\Upsilon{(4S)} \to B\kern 0.18em\overline{\kern -0.18em B}$ decays
collected with the
\mbox{\slshape B\kern-0.1em{\smaller A}\kern-0.1em B\kern-0.1em{\smaller A\kern-0.2em R}}
detector at SLAC.  We find no evidence for these decays, and we set
Bayesian 90\% confidence level upper limits on the branching fractions
${\cal B}(B^- \to \ensuremath{D^-_s}\xspace \phi)<1.9 \times 10^{-6}$
and ${\cal B}(B^- \to \ensuremath{D^{*-}_s}\xspace \phi)<1.2 \times
10^{-5}$.  These results are consistent with Standard Model
expectations.

%% file: pubboard/acknow_PRL.tex
We are grateful for the excellent luminosity and machine conditions
provided by our \pep2\ colleagues, 
and for the substantial dedicated effort from
the computing organizations that support \babar.
The collaborating institutions wish to thank 
SLAC for its support and kind hospitality. 
This work is supported by
DOE
and NSF (USA),
NSERC (Canada),
IHEP (China),
CEA and
CNRS-IN2P3
(France),
BMBF and DFG
(Germany),
INFN (Italy),
FOM (The Netherlands),
NFR (Norway),
MIST (Russia), and
PPARC (United Kingdom). 
Individuals have received support from CONACyT (Mexico), A.~P.~Sloan Foundation, 
Research Corporation,
and Alexander von Humboldt Foundation.

%% file: paper.bbl
\begin{thebibliography}{99}
\bibitem{voigtian} A Voigtian is the convolution of a Breit-Wigner lineshape
with a Gaussian resolution term.
\end{thebibliography}
